\documentclass[sigconf, screen, nonacm, timestamp=false]{acmart}

\AtBeginDocument{%
  \providecommand\BibTeX{{%
    \normalfont B\kern-0.5em{\scshape i\kern-0.25em b}\kern-0.8em\TeX}}}

\copyrightyear{2024}
\acmYear{2024}
\setcopyright{acmlicensed}
\acmConference[EASE 2024]{28th International Conference on Evaluation and Assessment in Software Engineering}{June 18--21, 2024}{Salerno, Italy}
\acmBooktitle{28th International Conference on Evaluation and Assessment in Software Engineering (EASE 2024), June 18--21, 2024, Salerno, Italy}
\acmDOI{10.1145/3661167.3661231}
\acmISBN{979-8-4007-1701-7/24/06}

\begin{document}

\title{Analyzing the Accessibility of GitHub Repositories for PyPI and NPM Libraries}

\author{Alexandros Tsakpinis}
\affiliation{%
  \institution{fortiss - Research Institute of the Free State of Bavaria}
  \city{Munich}
  \country{Germany}}
\email{tsakpinis@fortiss.org}

\author{Alexander Pretschner}
\affiliation{%
  \institution{Technical University of Munich}
  \city{Munich}
  \country{Germany}}
\email{alexander.pretschner@tum.de}

\begin{abstract}
    Industrial applications heavily rely on open-source software (OSS) libraries, which provide various benefits. But, they can also present a substantial risk if a vulnerability or attack arises and the community fails to promptly address the issue and release a fix due to inactivity.
    To be able to monitor the activities of such communities, a comprehensive list of repositories for the libraries of an ecosystem must be accessible. Based on these repositories, integrated libraries of an application can be monitored to observe whether they are adequately maintained.
    In this descriptive study, we analyze the accessibility of GitHub repositories for PyPI and NPM libraries. For all available libraries, we extract assigned repository URLs, direct dependencies and use the page rank algorithm to comprehensively analyze the ecosystems from a library and dependency chain perspective. For invalid repository URLs, we derive potential reasons.
    Both ecosystems show varying accessibility to GitHub repository URLs, depending on the page rank score of the analyzed libraries. For individual libraries, up to 73.8\% of PyPI and up to 69.4\% of NPM libraries have repository URLs. Within dependency chains, up to 80.1\% of PyPI libraries have URLs, while up to 81.1\% for NPM. That means, most libraries, especially the ones of increasing importance, can be monitored on GitHub. Among the most common reasons for invalid repository URLs is no URLs being assigned at all, which amounts up to 17.9\% for PyPI and  up to 39.6\% for NPM. Package maintainers should address this issue and update the repository information to enable monitoring of their libraries.
\end{abstract}

\ccsdesc[500]{Software and its engineering~Maintaining software}
\keywords{Maintenance Activities, OSS Libraries, Repository Mining}

\maketitle

\newpage
\section{Introduction}

Over time, Open Source Software (OSS) became a standard practice in the product life cycle due to commercial, engineering, and quality reasons \cite{ebert2008open}. Roughly 80\% of code in commercial products comprises OSS components, reflecting the widespread utilization in the software industry \cite{pittenger2016open}. Software libraries as key element of OSS play a key role in modern software development. Libraries, also called packages, enable the reuse of established, well-tested code for specific functionalities, reducing the need for coding from scratch \cite{bauer2012structured}. After a library has been included in a project, it can be defined as a dependency \cite{cox2019surviving}. OSS dependencies can be affected by vulnerabilities (e.g., Log4j\footnote{\url{https://logging.apache.org/log4j/2.x/}}) or supply-chain-attacks (e.g., npm colors\footnote{\url{https://snyk.io/de/blog/open-source-npm-packages-colors-faker/}}) representing significant challenges in the software industry \cite{decan2018impact}. If a vulnerability or an attack is identified, the library's community generally releases a fixed version soon after \cite{rahkema2022swiftdependencychecker}.

But, it can happen that support for a library is terminated or suspended resulting in the harmful situation of no maintenance activities being available \cite{bauer2012structured, raemaekers2011exploring}. Consequently, it is crucial that direct and transitive dependencies of an application are monitored by the application maintainers accordingly \cite{kula2014visualizing, tsakpinis2023analyzing}. Monitoring the maintenance activities of dependencies requires accessibility to corresponding repositories which are reachable via URLs linking to code management systems (CMS), such as GitHub. However, an exhaustive investigation into the accessibility of GitHub repository URLs for libraries of an ecosystem is currently missing.

In this work, we focus on GitHub as the CMS, given its ongoing popularity for hosting OSS libraries \cite{eghbal2020working}. Similar to other studies, we select PyPI and NPM as ecosystems due to their widespread use in modern software development and strong integration of libraries in their development practices \cite{decan2016topology, abdalkareem2020impact}. Thus, available GitHub repository URLs are collected for each library from their package managers. For libraries without valid URLs, possible reasons for their absence are documented. In this context, a valid URL is confirmed by an HTTP status code 200 and conforms syntactically to a standard GitHub URL scheme. Since libraries are heavily interconnected, they should not only be analyzed individually, but also include an analysis of their dependency chains. Thus, direct dependencies are collected for each library to create a dependency graph that is enriched with the available repository URL information. As the importance of individual libraries within their ecosystem can differ significantly, the dependency graph is used to calculate the importance of each library, as measured by its page rank score, and incorporate it in the analysis. This graph serves as the basis to gain insights into the accessibility of valid GitHub repository URLs from a library and dependency chain perspective.

\newpage
To enable the monitoring of maintenance activities for PyPI and NPM libraries on GitHub, insights into the accessibility of valid GitHub repository URLs are needed. The following research questions (RQ) aim to provide evidence for these insights:

\textbf{RQ1:} What is the ratio of libraries with valid GitHub repository URLs when evaluating the libraries of each ecosystem individually?

\textbf{RQ2:} What are possible reasons for missing valid GitHub repository URLs and what does the corresponding distribution look like?
 
\textbf{RQ3:} What is the ratio of libraries with valid GitHub repository URLs when focusing on dependency chains within each ecosystem? \newline

The descriptive analysis of the PyPI and NPM ecosystem shows mixed accessibility to valid GitHub URLs for individual libraries and libraries in dependency chains  depending on their page rank score. For individual libraries, up to 73.8\% of PyPI and up to 69.4\% of NPM libraries contain valid GitHub URLs. For libraries within dependency chains, up to 80.1\% of PyPI and up to 81.1\% of NPM libraries have valid URLs. As the accessibility is particularly positive for more important libraries, the maintenance activities for most of these libraries can be monitored. One of the main reasons for invalid URLs is the absence of any URL, which accounts for up to 17.9\% for PyPI and up to 39.6\% for NPM. To enable monitoring of maintenance activities for even more libraries and dependency chains, maintainers should update the repository information of their libraries in the package managers accordingly. Additionally, the analysis reveals insights into the complexity of the PyPI and NPM ecosystems. On average, there are around 3 direct and 37 transitive dependencies with URLs in PyPI, while 3 and 32 in NPM. Furthermore, PyPI shows maximum values of 270 direct and 667 transitive dependencies, while NPM holds 967 direct and 7780 transitive dependencies with URLs. Given the high complexity and the need to monitor the maintenance activities of dependencies, we recommend an automated monitoring approach in the future.
\section{Related Work}
In the following, related work is presented to highlight the importance of monitoring OSS libraries and to point out the differences of this work compared to other studies that analyze OSS ecosystems.

Recently, the US Cybersecurity and Infrastructure Agency (CISA) published a roadmap, focusing on the development of a framework for prioritizing OSS risks, also based on the level of maintenance. However, it lacks a specific definition, such as monitoring maintenance activities in a CMS. This roadmap highlights the importance of our research even for government-level organizations \cite{CISARoadmap}.

Numerous studies investigate package dependency networks for different ecosystems, focusing on security issues such as vulnerabilities \cite{dusing2022analyzing, alfadel2023empirical}, intrusion of malicious code \cite{guo2023empirical} and evolution of ecosystem robustness \cite{hafner2021node}. Another well-researched area is package versioning, which includes updating strategies \cite{javan2023dependency}, configuration issues \cite{peng2023less}, dependency conflicts \cite{wang2020watchman}, smells \cite{cao2022towards} and errors \cite{mukherjee2021fixing}. In addition, the characteristics of highly selected NPM packages \cite{mujahid2023characteristics}, the determinants of sustainable OSS Python projects \cite{valiev2018ecosystem} as well as the usage of trivial packages in PyPI and NPM \cite{abdalkareem2020impact} are investigated. All these studies analyze various aspects of ecosystems but none has investigated the accessibility of valid GitHub repository URLs for their analysis, so our study aims to fill this gap.

Unlike the studies mentioned before focusing on specific aspects of an ecosystem, some studies examine broader characteristics such as dependencies \cite{bommarito2019empirical, decan2016topology}, with a focus on transitive dependencies \cite{decan2019empirical, kikas2017structure}. While numerous studies analyze libraries and their dependencies for different ecosystems, none has specifically addressed the accessibility of valid repository URLs for libraries within dependency chains. Thus, our research also seeks to close this gap.
\section{Research Method}
In the following, our research method for this descriptive study is explained in more detail, with a focus on how the data collection and data analysis have been conducted. The primary goal is to gather evidence to assess the accessibility of GitHub repository URLs enabling a CMS-based monitoring of maintenance activities.

\subsection{Data Collection}
\label{sec:data_collection}
To assess the accessibility of valid GitHub repository URLs within an ecosystem, we automatically collect assigned URLs for each library using different Python scripts. Therefore, a comprehensive list of available libraries is required which is provided by corresponding endpoints\footnote{PyPI: \url{https://pypi.org/simple/}\label{all_endpoint_pypi}}\textsuperscript{,}\footnote{NPM: \url{https://skimdb.npmjs.com/registry/_all_docs/}\label{all_endpoint_npm}}. For details about each library, such as assigned URLs and dependency information, additional endpoints\footnote{PyPI: \url{https://pypi.org/pypi/<package\_name>/json}\label{detailed_endpoint_pypi}}\textsuperscript{,}\footnote{\label{detailed_endpoint_npm}NPM: \url{https://skimdb.npmjs.com/<package_name>}} are consumed. As dependency information, we collect all direct and optional dependencies for the latest published library version to create a dependency graph fully capturing the dependency tree without limiting the transitive depth. For each assigned GitHub URL, automated validity checks are performed, saving valid URLs or documenting one of the following reasons for invalid URLs:

\begin{itemize}
    \item No URL assigned at all 
    \item 404 status code returned by ecosystem endpoints\textsuperscript{\ref{detailed_endpoint_pypi}}\textsuperscript{,}\textsuperscript{\ref{detailed_endpoint_npm}}
    \item 404 or other GitHub related status codes 
    \item Use of private or unknown CMS
    \item Use of GitLab or BitBucket as CMS
    \item Malformed CMS URLs 
\end{itemize}

The data collection results in a JSON file for each ecosystems linking library names with their direct dependencies, and either a repository URL or a potential reason for the absence of a valid URL.

\subsection{Data Analysis}
In our data analysis, we use standard terms such as nodes, dependency chains and dependency graph. Each library represents a node in the dependency graph and is connected to its direct dependencies by dependency chains. Together, this network of nodes and dependency chains forms the dependency graph representing an ecosystem. The data from Section \ref{sec:data_collection} serve as the basis for various analyses providing insights from two perspectives. First, all libraries should be represented as individual nodes without considering dependency chains providing a library perspective on the ecosystems. Second, the data should be analyzed from a dependency chain perspective, focusing on all dependency chains originating from existing nodes within the dependency graph. Therefore, the following metrics are calculated for the library perspective:

\begin{itemize}
    \item Ratio of packages with valid repository URLs
    \item Distribution of reasons for invalid URLs (see Section \ref{sec:data_collection})
\end{itemize}

These are the metrics for the dependency chain perspective:

\begin{itemize}
    \item Ratio of libraries in dependency chains having URLs
    \item Average and maximum number of direct and transitive dependencies with URLs
\end{itemize}

Analyzing nodes in isolation provides a basic understanding of individual libraries and focuses on their standalone properties. In contrast, looking at dependency chains provides a more comprehensive view on the structure of the ecosystem leading to insights that are not apparent when looking at individual libraries. 

Note that the metrics for both perspectives are calculated for different subsets of the dependency graph. As a measure for defining subsets, we use the page rank algorithm inspired by \citeauthor{mujahid2021toward}, where each node within a dependency graph is assigned an importance value based on its incoming and outgoing dependency chains \cite{mujahid2021toward}. This approach of categorizing libraries according to their importance enables a more differentiated understanding and improves the analysis of libraries using different levels of importance.

Since the importance value of the page rank algorithm is largely incorporated in the data analysis, we also conduct a correlation analysis to determine whether any of the calculated variables show a correlation with the importance value.
\section{Results}
The data collection was carried out on February 16, 2024 resulting in 514,535 PyPI and 2,679,007 NPM libraries. When analyzing the dependencies for each library, we neglect libraries that can not be found in the initial ecosystem's library list\textsuperscript{\ref{all_endpoint_pypi}}\textsuperscript{,}\textsuperscript{\ref{all_endpoint_npm}}. In total, we were not able to find 0.3\% of PyPI and 0.8\% of NPM libraries from a node and each 0.2\% from a dependency chain perspective. Also we exclude 3.0\% of PyPI and 33.9\% of NPM libraries which return a status code (mostly 404) when querying the detailed ecosystem endpoints\textsuperscript{\ref{detailed_endpoint_pypi}}\textsuperscript{,}\textsuperscript{\ref{detailed_endpoint_npm}}. These errors arise when the package name retrieved from the comprehensive library list endpoint\textsuperscript{\ref{all_endpoint_pypi}}\textsuperscript{,}\textsuperscript{\ref{all_endpoint_npm}} could not be found at the detailed endpoint\textsuperscript{\ref{detailed_endpoint_pypi}}\textsuperscript{,}\textsuperscript{\ref{detailed_endpoint_npm}}, indicating inconsistencies between the two endpoints which should be fixed by the ecosystems. The final dataset has 499,274 PyPI libraries with 1,068,768 dependency chains, while 1,770,463 NPM libraries with 5,408,141 dependency chains.

The analysis focuses on libraries with increasing importance measured by the page rank score, as these nodes have a greater influence within the dependency tree and are used more frequently. The x-axis of the shown figures defines the subset on which the characteristics are calculated, e.g., $x=10$ specifies a subset of more important libraries, which contains the top 10\% of libraries ranked by page rank score, while $x=100$ describes the top 100\% of libraries, which corresponds to the entire data set also containing less important libraries. Consequently, the results are often described as value ranges relating to the different subsets. If not stated otherwise, the observations mentioned in the following apply to both ecosystems, and the term URL refers to a valid GitHub repository URL. 

\subsection{RQ1+RQ2: Library perspective}
\label{sec:node_perspective}

\begin{figure}[h]
    \centering
    \includegraphics[width=0.47\textwidth]{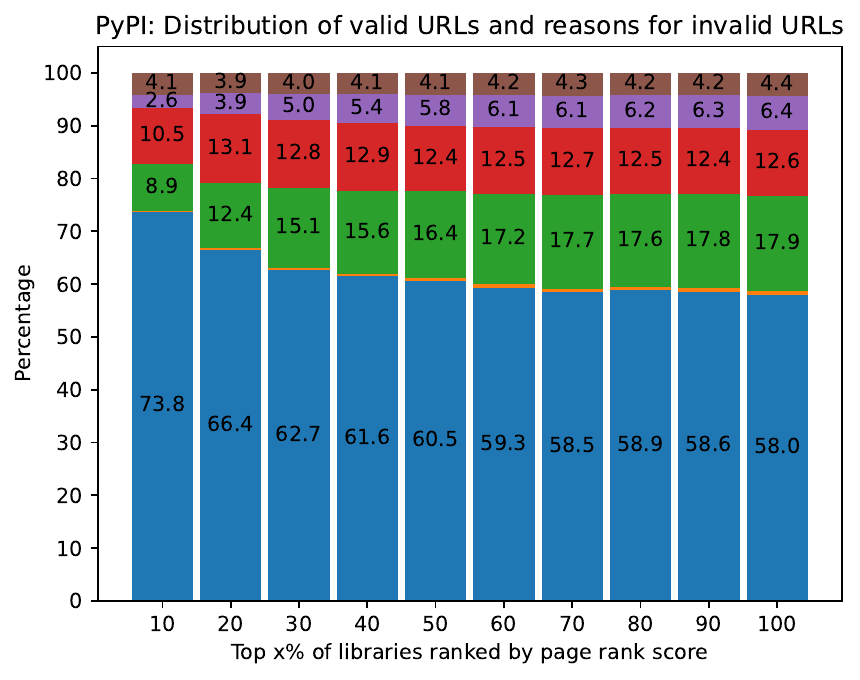}
    \hfill
    \includegraphics[width=0.47\textwidth]{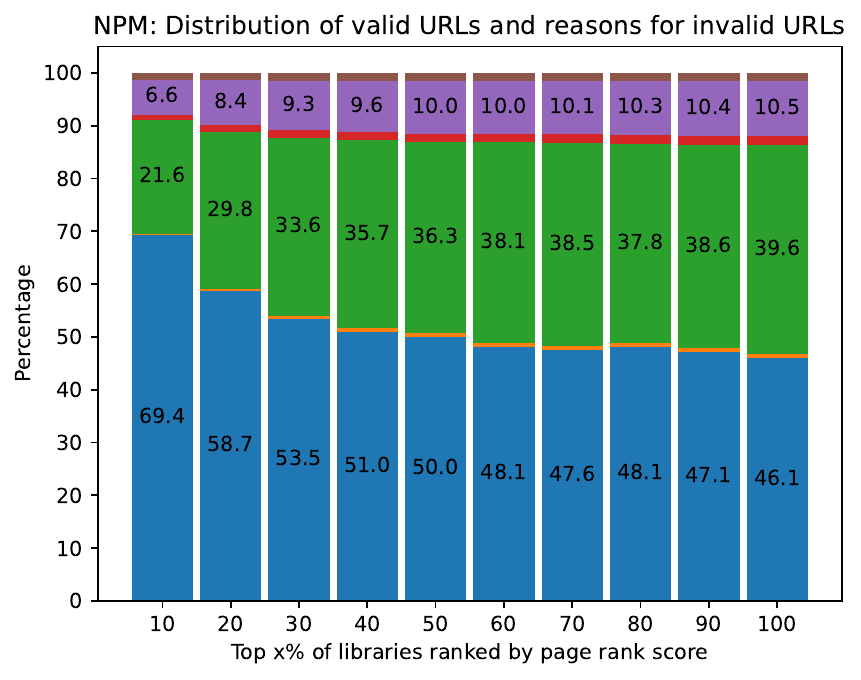}
    \hfill
    \includegraphics[width=0.47\textwidth]{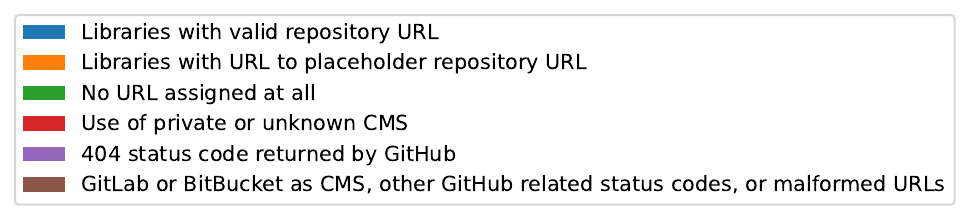}
    \caption{Distribution of URLs and reasons for invalid URLs}
    \label{fig:results_node_perspective}
\end{figure}

Figure \ref{fig:results_node_perspective} presents the results of the data analysis of how the characteristics of the nodes in the dependency graph behave independently without looking at dependency chains. In particular, the distribution of valid GitHub repository URLs and various reasons for invalid URLs are computed, as introduced in Section \ref{sec:data_collection}.

Valid URLs are further divided into "real" valid and placeholder URLs (orange), which are used for sample projects\footnote{\url{https://github.com/pypa/sampleproject}} in PyPI, or for insecure\footnote{\url{https://github.com/npm/security-holder}} or deprecated\footnote{\url{https://github.com/npm/deprecate-holder}} packages in NPM. The overall ratio of placeholder URLs is 0.7\% for both PyPI and NPM. Even if these ratios are neglectable, developers should avoid such libraries, as these repositories do not reflect the actual code base of the library, which prevents monitoring of maintenance activities in a CMS.

Figure \ref{fig:results_node_perspective} shows, that more important libraries tend to have an increasing percentage of URLs (blue), with values ranging from 58.0\% to 73.8\% for PyPI and from 46.1\% to 69.4\% for NPM. Consequently, the accessibility from a node perspective is not too promising even for libraries with increasing importance, evidenced by maximum values of 73.8\% for PyPI and 69.4\% for NPM. Additionally, there are some severe URL gaps for less important NPM libraries preventing the monitoring of maintenance activities in a CMS. This observation can be seen by the increase in libraries with URLs when moving from $x=100 $ to $x=10$, where the values increase by 50.5\%, while less important libraries are more and more neglected.

A significant fraction of packages, especially visible for NPM, have no URL assigned at all (green), which accounts for 8.9\%-17.9\% for PyPI and 21.6\%-39.6\% for NPM. The ratios of both ecosystems decrease with increasing library importance. Such occurrences of missing URLs might be an indicator of poor maintenance practices when updating package metadata. Maintainers could take these findings to review their packages and ensure that the repository URLs are valid and reflect the current state of their codebase. Another interpretation may be that some libraries are closed-source and do not allow access to their repositories on purpose.

There are also GitHub-related problems with status code 404 (purple) representing a syntactically valid GitHub repository URL that cannot be found at this location. With increasing importance, PyPI's ratios decrease from 6.4\% to 2.7\%, while NPM's ratios decrease from 10.5\% to 6.6\%. The widespread occurrence of such errors indicates a need for improvement in the package update processes, as numerous repositories have been moved or deleted without the package information being updated accordingly.

PyPI shows small percentages for URL-related errors such as malformed URLs, various GitHub related status codes, or GitLab or Bitbucket as CMS (brown) consistently having values below 4.4\%. But, between 10.5\% and 13.1\% are assigned to unknown CMS providers (red) meaning that occasionally, links lead to not widely known CMS providers, such as those not affiliated with GitHub, GitLab, or Bitbucket, or even to private homepages instead of publicly accessible CMSs. We recommend that package maintainers replace these not publicly accessible CMSs, as they hinder the monitoring of maintenance activities. For the NPM ecosystem, errors related to URLs (brown) or unknown CMS providers rarely occur (red) with values consistently below 1.7\%. In particular, the absence of the latter category suggests that more publicly accessible CMS providers are linked in NPM, which is a positive sign regarding the accessibility of valid GitHub repository URLs.

\subsection{RQ3: Dependency chain perspective}
\label{sec:dependency_chain_perspective}

\begin{figure}[h]
    \centering
    \includegraphics[width=0.47\textwidth]{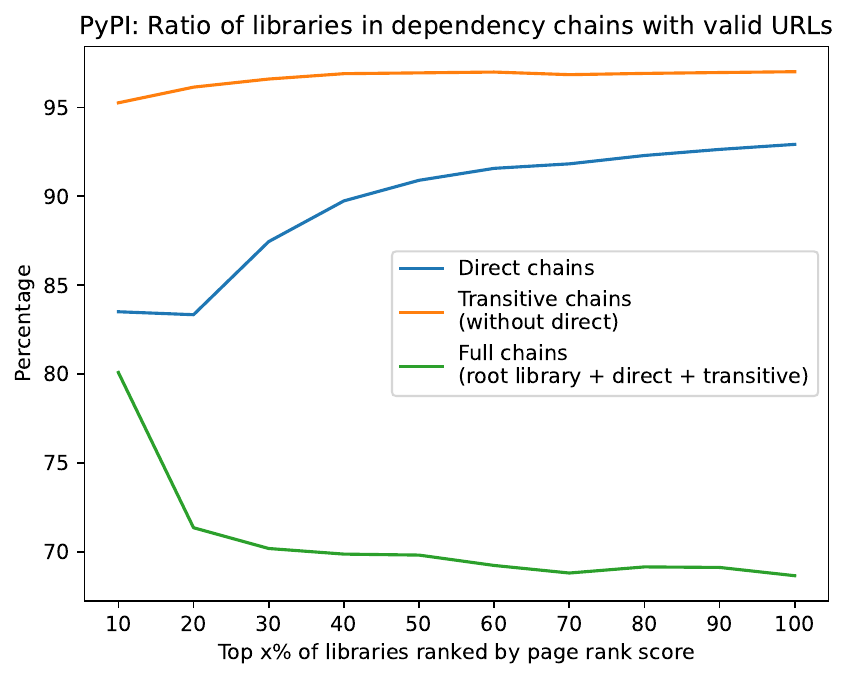}
    \hfill
    \includegraphics[width=0.47\textwidth]{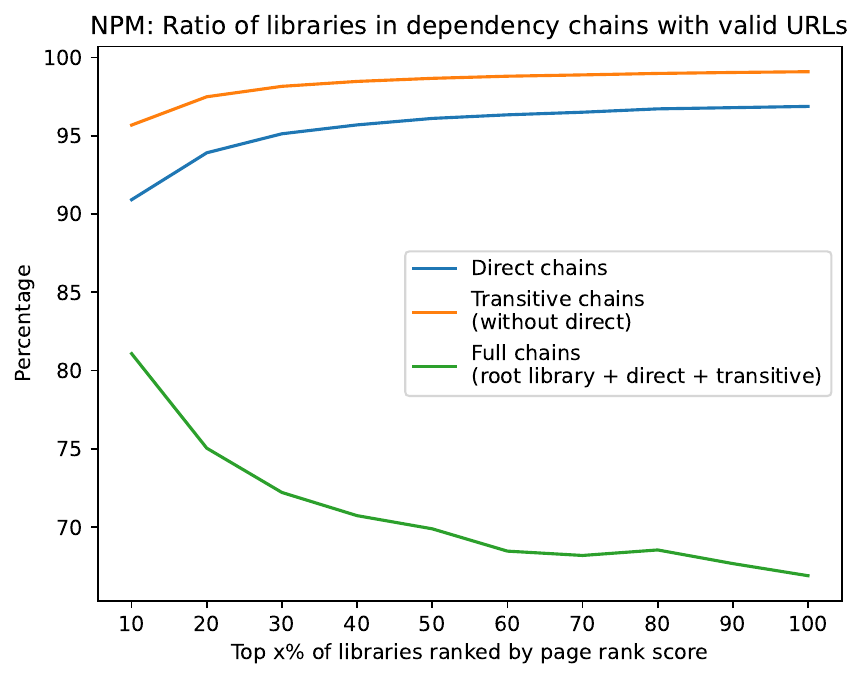}
    \caption{Ratio of libraries in chains having URLs}
    \label{fig:percentage_nodes_havin_valid_url}
\end{figure}

In the following, different characteristics of the libraries including their dependency chains are analyzed. First, Figure \ref{fig:percentage_nodes_havin_valid_url} analyzes direct, transitive and full chains regarding the ratio of libraries within the dependency chains to which an URL is assigned describing a measure for the accessibility of URLs. In this context, \textit{direct chains} refer to dependencies that are explicitly chosen by a library. \textit{Transitive chains} include all dependencies that are transitively connected but cannot be directly influenced. A \textit{full chain} represents the sum of direct and transitive dependencies, plus the characteristics of the root library itself.

Figure \ref{fig:percentage_nodes_havin_valid_url} shows that with increasing importance, between 68.7\% and 80.1\% of the libraries within full dependency chains have URLs for PyPI, while between 66.9\% and 81.1\% for NPM. Within direct dependency chains, between 83.3\% and 92.9\% of PyPI libraries and between 90.9\% and 96.9\% of NPM libraries have access to URLs. In transitive chains, between 95.3\% and 97.0\% of PyPI libraries and 95.7\% to 99.1\% of NPM libraries can access URLs. Overall, the high ratios, especially for more important libraries, are a good indication that most libraries within the dependency chains provide access to valid GitHub URLs enabling monitoring of maintenance activities. Interestingly, the ratios for direct and transitive dependency chains are always higher than for full ones. This means that, the root libraries often lack URLs, but their direct and transitive dependencies have URLs assigned. Thus, most missing URLs in libraries within full dependency chains can be corrected by the package maintainers themselves, as they are in control over the repository information of the library uploaded to the ecosystems.

\begin{figure}[h]
    \centering
    \includegraphics[width=0.47\textwidth]{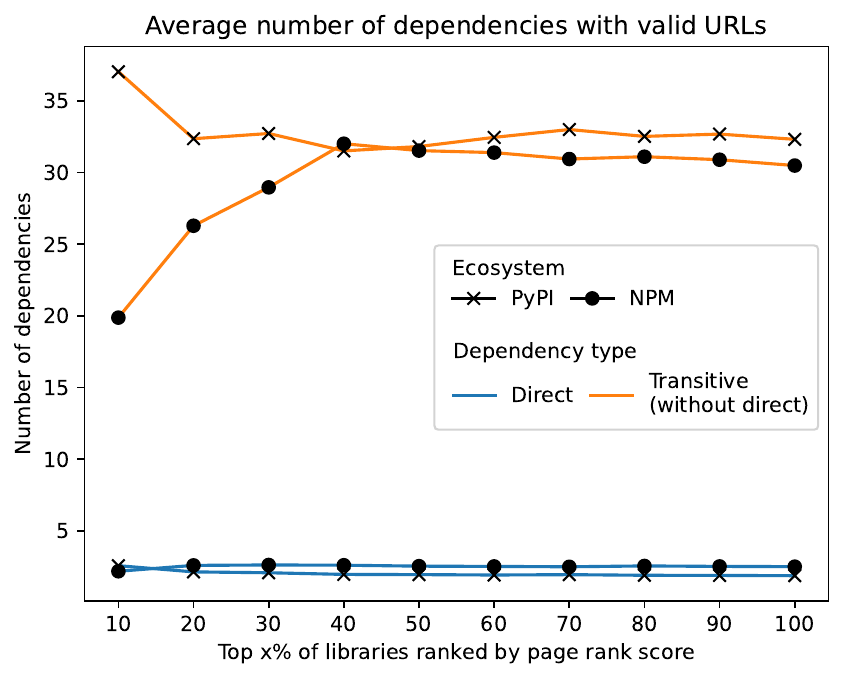}
    \caption{Direct and transitive dependencies with URLs}
    \label{fig:chains_with_and_without_URLs}
\end{figure}

Second, the average number of direct and transitive dependencies with URLs is calculated indicating the complexity of each ecosystem. Figure \ref{fig:chains_with_and_without_URLs} shows the average number of direct and transitive dependencies with URLs having values up to around 3 and 37 for PyPI, and up to 3 and 32 for NPM. Note, that direct dependencies are not included in the transitive ones. The large number of transitive dependencies with URLs multiplied with the number of direct ones shows the high complexity of the dependency graphs and provides evidence that manual monitoring is not feasible describing the need for an automated monitoring approach in the future.

Third, the maximum number of direct and transitive dependencies with URLs is measured for each ecosystem which additionally provides evidence for the need for an automated monitoring approach in the future. Note, that these observations are not shown in any Figure. PyPI's maximum values are 270 direct and 667 transitive dependencies with URLs, while NPM's maxima are 967 and 7780.

\subsection{Correlation analysis}
To evaluate different hypotheses regarding the importance value of the page rank algorithm, which plays a central role in this study, we conduct various correlation analyses between different features of a library and its importance value. Therefore, we use the Pearson coefficient in order to find linear relationships \cite{cleophas2018bayesian}.

In particular, we carry out three correlation analyses which are motivated by the results presented in Figure \ref{fig:results_node_perspective}, \ref{fig:percentage_nodes_havin_valid_url} and \ref{fig:chains_with_and_without_URLs}. Intuitively, our hypotheses are that more important libraries would have a higher chance of being assigned an URL (Figure \ref{fig:results_node_perspective}), would contain higher ratios of libraries in direct, transitive and full dependency chains with URLs (Figure \ref{fig:percentage_nodes_havin_valid_url}), and would exhibit a higher number of direct and transitive dependencies with URLs (Figure \ref{fig:chains_with_and_without_URLs}). Surprisingly, all our hypotheses about expected correlations could not be confirmed with Pearson coefficients ranging from -0.0029 to 0.0075 for PyPI, and values between -0.0023 and 0.011 for NPM. These results indicate that there is almost no relationship between the analyzed features as the values are close to zero. 

The weak correlations can be explained by the study's focus on libraries with increasing importance, while less important libraries are neglected for deeper analysis. Further investigations show that less important libraries receive more URLs than expected, and thus relativize the positive results of more important libraries, which finally leads to such low Pearson coefficients.
\section{Discussion}
We investigated the accessibility of valid GitHub repository URLs from a node and dependency chain perspective by using different subsets of libraries varying their importance to finally gain a deeper understanding of the ecosystems. In this context, the results of the most important libraries are of greater relevance as they play a central role in their ecosystems and exert a significant influence.

From a library perspective, the results for PyPI and NPM libraries are not overly promising regarding the accessibility of valid GitHub repository URLs, not even for their most important libraries. This applies particularly for less important NPM libraries where a significant portion of URLs is missing. For both ecosystems we identified clear areas where improvements could be made. This may include better tooling for repository URL verification, automated checks for link validity, and more robust community practices for updating package metadata. In addition, library ecosystems should focus on eliminating inconsistencies between the comprehensive package name lists and the detailed package endpoints they provide. 

But, even more important than the library perspective is the dependency chain perspective when assessing the accessibility of URLs. This is because the ratio of how many libraries within a dependency chain have an URL assigned provides a more realistic view since libraries rarely appear alone without having dependencies themselves. The ratio of libraries with valid GitHub repository URLs within direct, transitive and full dependency chains shows positive signs for both ecosystems, especially for libraries with increasing importance. Interestingly, most missing URLs can be found in root libraries, which could be immediately corrected by the package maintainers themselves. The results on existing dependency chains also reflect the high complexity of these ecosystems, given by the large average and maximum number of direct and transitive dependencies with valid GitHub repository URLs, making an automated monitoring approach for maintenance activities indispensable in the future. 
\section{Threats to Validity}
Threats to validity are divided into the following four aspects \cite{runeson2009guidelines}: 

\textbf{Construct validity:}
For the primary study objective of analyzing the accessibility of valid GitHub repository URLs, there is no threat to construct validity, as the URLs are checked against a standard GitHub URL scheme and an HTTP status code 200 without any interpretation. A minor threat to construct validity arises when interpreting and deriving reasons for invalid URLs. This threat is mitigated by deriving reasons only from objective patterns (e.g., status codes, CMS provider names, URL format) found in the dataset. Another threat arises from our approach to resolving dependencies for a particular library. We disregarded version information for any dependency, consistently opting for the latest version. This may not accurately reflect the entire ecosystem's dynamics. Nevertheless, we contend that the resulting discrepancies in the calculated page rank score of a library are negligible.

\textbf{Internal validity:} 
No threat to internal validity, as we describe ecosystem characteristics without investigating causal relations.

\textbf{External validity:} 
We focus on GitHub as CMS, and PyPI and NPM as ecosystems. Although the results may not generalize to other CMS or ecosystems, the applied research method can be transferred to any publicly crawlable ecosystem where libraries can include links, such as CMS references, in their metadata.

\textbf{Reliability:} 
While all data are collected from publicly accessible sources, which theoretically ensures reproducibility of the results, a threat to reliability still arises due to the nature of the ecosystems. The collected data reflect a specific timestamp from the collection day, posing a challenge for reproducibility as these ecosystems typically lack built-in time-travel functionality. To ensure complete reproducibility, the collected data and analysis code are published on figshare \cite{tsakpinis_pretschner_2024}.
Another threat is the frequent emergence or modification of libraries posing a risk of gaps or outdated information in our dataset. To address this threat, the study must be repeated.

\section{Conclusion and Future Work}
The paper analyzed the accessibility of valid GitHub repository URLs for the PyPI and NPM ecosystems. This is necessary for monitoring the maintenance activities of OSS libraries within their respective CMS, which enables the identification of poorly maintained libraries representing a potential security risk. We analyzed the PyPI and NPM ecosystem from a library and a dependency chain perspective providing a comprehensive view on the accessibility of valid GitHub repository URLs. Particularly, the high ratios of libraries with URLs in full dependency chains (PyPI: up to 80.1\%, NPM: up to 81.1\%) serve as proof that the maintenance activities of most libraries in dependency chains can be monitored on GitHub, which is especially true for libraries of increasing importance.

Future work should explore the reasons for missing, outdated or invalid URLs to understand why maintainers neglect such maintenance practices. Additionally, we aim to periodically repeat this study to observe ecosystem evolutions regarding GitHub repository URL accessibility. Given the high number of direct and transitive dependencies with GitHub repository URLs, we also plan to investigate what concrete metrics capture the maintenance activities of OSS libraries on GitHub, which can be used to develop an automated monitoring approach identifying poorly maintained libraries.

\bibliographystyle{ACM-Reference-Format}
\bibliography{bibliography}

\appendix

\end{document}